# Temporal Clustering in Dynamic Networks with Tensor Decomposition

Kun Tu    Bruno Ribeiro    Ananthram Swami    Don Towsley

**Abstract**

Dynamic networks are increasingly being usedd to model real world datasets. A challenging task in their analysis is to detect and characterize clusters. It is useful for analyzing real-world data such as detecting evolving communities in networks. We propose a temporal clustering framework based on a set of network generative models to address this problem. We use PARAFAC decomposition to learn network models from datasets. We then use $K$-means for clustering, the Silhouette criterion to determine the number of clusters, and a similarity score to order the clusters and retain the significant ones. In order to address the time-dependent aspect of these clusters, we propose a segmentation algorithm to detect their formations, dissolutions and lifetimes. Synthetic networks with ground truth and real-world datasets are used to test our method against state-of-the-art, and the results show that our method has better performance in clustering and lifetime detection than previous methods.

Synthetic data and experiment codes can be obtained from github
( https://github.com/submissionCode/KDD2017)

# Contents

# 1 Introduction

Clustering in dynamic networks is an important task in network science for knowledge discovery. With a correctly defined similarity/distance metric, clusters suggest "meaningful" communities where nodes in the graph are more densely connected. For real-world social networks or communication networks, these clusters usually represent groups of close friends or frequent contacts. Identifying such clusters in a time varying network helps to understand the network structure. This allows further to infer the relationships among nodes and to better understand the network dynamics.

Given a dynamic network represented as network snapshots, the task of temporal clustering focuses on identifying clusters of nodes and detecting their lifetimes. This task is challenging because it is usually difficult to decide the appropriate number for clusters, whose value can significantly affects clustering results. Moreover, clusters are evolving, making lifetime detection difficult.

Related work mainly focuses on two different approaches to solve this task. The first approach focuses on identifying clusters with a set of snapshots and obtains their lifetimes based on the time steps when they are detected. Since there could be many clusters generated from all snapshots, previous work such as [7] only maintains those most frequently appearing clusters. Evolutionary clustering (EC) [6] applies $K$-means clustering to a similarity matrix generated from the current snapshot and the clustering result of previous snapshots. Researchers have replaced the $K$-means clustering with other clustering methods or proposed new ways to generate simialarity matrices [26, 16, 3, 14, 28] to improve EC. However, these methods use local information (more specifically, a similarity matrix generated from a small subset of snapshots) to identify clusters. When snapshots are sparse and contain too few edges, the similarity matrices provide little information for clustering, and the detected clusters can be small and fail to correspond to meaningful clusters. Moreover, nodes are usually assigned to only one cluster in a time step, while in real-world social networks or email networks, a person may be part of multiple communities at the same time.

The other approach designs a model of a dynamic network with evolving latent structures and uses global information (all snapshots) to learn those structures. Evolving stochastic block models (SBM) [9, 30, 15] are proposed for a dynamic network where each node has a mixed membership of communities defined by the model. A probability model is applied to learn the model. However, since these methods focus on computing the memberships, the formation, dissolution and lifetime of a community remains unknown. Tensor decomposition based methods such as [10, 17, 21] model a network as a three mode tensor and apply low-rank tensor factorization to obtain $R$ components. Each component consists of three vectors named "loading vectors". Two of the loading vectors relate to nodes and are used to generate a community with binary classification. The other loading vector contains temporal information for tracking the community lifetime. However, previous analysis with tensor decomposition fails to provide a good model for the dynamic network; more precisely, the physical interpretations of the vectors related to nodes and the time are unclear. As a result, they inaccurately determine the lifetime of a community when network snapshots are sparse and contain few edges. They also fail to provide a way to accurately calculate the lifetime of communities. Some methods merge snapshots to analyze data at a large time granularity, but this can result in inaccurate lifetime detection for a cluster because of the loss in details of the change in a cluster.

To overcome the issues caused by sparse snapshots, we propose a temporal clustering framework based on a set of latent network generative models and PARAFAC decomposition. Our model of a dynamic network contains multiple generative models, each of them corresponds to one or more several clusters. Given a time interval, edges between nodes in a cluster are generated at a specific rate. We assume the edge-generating rate changes at a speed slower than the network structure and use it to detect the formation, dissolution and lifetime of the clusters. We will formalize our model in the next section.

The major contributions of our work include:

- We provide a temporal clustring framework to detect evolving communities in a dynamic network using $K$-means and Silhouette criterion. Our model allows a node to belong to multiple communities and improves the accuracy of ground truth community detection. Its performance is robust to changes in time granularity and the sparsity/density of eges in snapshots.

- We model the rate of edge generation at a cluster level. This allows our model to deal with different time granularities and to have better performance in lifetime detection, especially when network snapshots are too sparse to provide structure information for clustering.

The organization of our paper is as follows: Section 2 formalizes our temporal clustering model and problem. Section 3 explains the steps in temporal clustering to learn the latent structures and detect communities and their lifetime. Section 4 reports experiments with both synthetic data and real data to test our method.

## 2 Temporal Clustering Model

### 2.1 Generative Model for Dynamic Network

We model a dynamic network $G = \{G_t(V, E_t) | t = 1, \cdots, T\}$ as a mixture of $R$ generative models $\{X^{(r)}\}_{r=1}^{R}$, where $G_t$ is the network snapshot at time $t$, $V$ is a fixed set of nodes and $E_t$ is the set of edges at $t$. $X^{(r)}$ contains the same set of nodes $V$, for $r = 1, \cdots, R$. We assume that the probability that a node $i \in V$ belongs to $X^{(r)}$ follows a Bernoulli distribution, denoted as $P(i \in X^{(r)}) = a_{ir}$. At time step $t$, the $r$-th generative model $X^{(r)}$ generates and edge between two nodes $i, j$ with probability $a_{ir}a_{jr}\lambda^{(r)}(t)$, where $\lambda^{(r)}(t)$ is defined as an *edge-generating rate*. $(a_{ir}a_{jr}\lambda^{(r)}(t)$ is interpreted as expected number of edges genereated between $i$ and $j$ in $X^{(r)}$ at $t$). $\lambda^{(r)}(t)$ changes throughout time and can be modeled as a time series or a specific function (Details are provided in Section 3.4).

Our model allows a network to have self-loops, which is defined as a edge connects a node to itself. If a self-loop is not allowed, our model can be easily extended by adding a binary weight $W_{ij}$ for calculating the edge-generating rate between a node pair $i, j$, where $W_{ij} = 1$ if $i \neq j$ and $W_{ij} = 0$ otherwise. (Please refer to supplemental materials for details at the end of this paper). In the rest of the paper, we explain our model assuming that self-loop is allowed.

### 2.2 Problem Formulation

we use a three-mode tensor $\mathcal{X}$ to represent a dynamic graph $G = \{G_t(V, E_t)\}_{t=1}^{T}$, where $\mathcal{X}_{ijt}$ represents the number of edges observed between node $i$ and $j$ at $t$. Our model approximates $\mathcal{X}_{ijt}$ with $\sum_{r=1}^{R} a_{ir}a_{jr}\lambda^{(r)}(t)$

The temporal clustering problem is to find $R$ network generative models $X^{(r)}$, their edge-generating rates $\{\lambda^{(r)}(t)\}$, and the probability that node $i$ belongs to the generative model, $a_{ir}$, for $r = 1, \cdots, R$, $t = 1, \cdots, T$, and $i = 1, \ldots, |V|$, so as to minimize :

$$\begin{aligned} \min & \sum_{i,j \in V} \sum_t (\mathcal{X}_{ijt} - \sum_{r=1}^{R} a_{ir}a_{jr}\lambda^{(r)}(t))^2 \\ \text{s.t} \quad & 0 \leq a_{ir} \leq 1; \text{ for } i = 1, \ldots, |V| \\ & \lambda^{(r)}(t) \geq 0 \end{aligned} \quad (1)$$

The problem applies for networks that allow self-loops. If self-loops are not allowed, another constraint, $i \neq j$ can be added to the objective.

### 2.3 Detecting Clusters with Generative Models

We allow $X^{(r)}$ to contain one or more clusters. The similarity between two nodes, $i$ and $j$ in $X^{(r)}$, is defined as:

$$s_{i,j}^{(r)} = a_{ir}a_{jr} \quad (2)$$

Clustering methods such as $K$-means algorithm and spectral clustering return disjoint clusters. Let $C_m^{(r)}$ be the $m$-th cluster obtained from generative model $X^{(r)}$, when our model applies those methods, $|C_m^{(r)} \cap C_n^{(r)}| = 0$. However, clusters across different generative models are allowed to overlap: $|C_m^{(r)} \cap C_n^{(s)}| \geq 0$, for $r \neq s$.

The approach to solving our problem consists of the following steps: (1) learning the $R$ generative models, (2) clustering, (3) cluster ranking, and (4) lifespan identification. We provide more details on each step in the next section.

## 3 Learning and Tracking Dynamic Communities

### 3.1 Learning Generative Model via PARAFAC Decomposition

Our temporal clustering framework learns $R$ generative models $X^{(r)}$ from the tensor $\mathcal{X}$ and uses these models to detect clusters. These models include information from all snapshots and avoid the problem that a sparse snapshot provides too little structure information for clustering. To obtain generative models that satisfy Eq (1), we apply PARAFAC decomposition.

A PARAFAC model decomposes a tensor into multiple components via a high-order singular value decomposition (HOSVD). It has been applied in many fields and a good reference is the survey from [4]. Given a three-mode tensor $\mathcal{X}$, a rank-$R$ PARAFAC decomposes $\mathcal{X}$ into $R$ components, each of which consists of a scalar $\lambda_r$ and a rank-1 tensor produced by the outer product of 3 column vectors $A_r \otimes B_r \otimes \mathcal{T}_r$, with the objective of minimizing the Frobenius norm of the error:

$$f(\{\lambda_r, A_r, B_r, \mathcal{T}_r\}_{r=1}^R) = \|\mathcal{X} - \hat{\mathcal{X}}\|_{\mathbf{F}} = \sqrt{\sum_{i,j,k}(\mathcal{X}_{ijk} - \hat{\mathcal{X}}_{ijk})^2} \quad (3)$$

where $\hat{\mathcal{X}} = \sum_{r=1}^R \lambda_r A_r \otimes B_r \otimes \mathcal{T}_r$, $A = [A_1, \cdots, A_R], B = [B_1, \cdots, B_R]$ and $\mathcal{T} = [\mathcal{T}_1, \cdots, \mathcal{T}_R]$ are call loading matrices. The PARAFAC decomposition model can be written as:

$$\mathcal{X}_{ijk} \approx \sum_{r=1}^R \lambda_r a_{ir} b_{jr} t_{kr} \quad (4)$$

where $a_{ir}$ is the $i$-th element of $A_r$, $b_{jr}$ is the $j$-th element of $B_r$ and $t_{kr}$ is the $k$-th element of $\mathcal{T}_r$.

Since the objective Eq (1) of our model is similar to that of PARAFAC, Eq (3), we add two constraints to Eq (3) and apply PARAFAC decomposition to learn our generative models: (1) the non-negative constraint $\lambda_r, a_{ir}, b_{jr}, t_{kr} \geq 0$ and (2) the symmetric constraint $A_r = B_r$ since the snapshots are symmetric. We also normalize $a_{ir}$ in interval $[0, 1]$, as $\max_i(a_{ir}) = 1$, then represent the edge-generating rate $\lambda^{(r)}(k)$ as a piecewise linear segmentation on the sequence of $T$ samples $\vec{\lambda}_r = [\lambda_r t_{kr}]$ for $k = 1, \cdots, T$.

In the rest of the paper, we use $\hat{X}^{(r)} = \{A_r, \vec{\lambda}_r\}$ as the $r$-th generative model learned from PARAFAC, where $a_{ir} \in A_r$ is the probability that node $i$ belongs to $\hat{X}^{(r)}$, $\vec{\lambda}_r = [\lambda_{1,r}, \ldots, \lambda_{T,r}]$ is a sequence of $T$ samples for the edge-generating rate $\lambda^{(r)}(t)$. Other notations are as defined in Section 2.

There are many approximate algorithms for PARAFAC decomposition [2, 1, 29, 24]. We apply the alternating least squares algorithm (ALS) [24] because of its good performance and fast speed on average. It applies a gradient descent method to optimize Eq (3) iteratively. At an iteration, each loading matrix is updated while other matrices are fixed.

Note that all current methods for learning latent structures require setting the number of structures either manually or via criteria such as the elbow rule[13] and modularity gain[18]. For PARAFAC decomposition, we select $R$ according to *core consistency* [5]. However, the rank $R$ is not necessarily consistent with the data. Moreover, PARAFAC decomposition requires large computation power, so that $R$ has to be small for large datasets. As a result, errors may occur in the components, making them unsuitable for analysis. For example, in an experiment by Gauvin et al. [10], one of the components from PARAFAC actually contains two independent communities, but they methods groups the two communities as one. We deals with this issue by performing clustering on $X^{(r)}$ and propose a ranking score to retain the significant clusters.

### 3.2 Clustering on Generative Models

After obtaining $R$ latent generative models, we perform clustering on each model $\hat{X}^{(r)}$ using the similarity defined in Eq (2). There are many clustering algorithms such as $K$-means and spectral clustering[20], as well as community detection algorithm such as Girvan−Newman algorithm [19]. We choose $K$-means algorithm with Silhouette criterion to determine the best number of clusters. There are mainly two reasons. First,there is no advantages in performance of advanced algorithms over $K$-means clustering. In $\hat{X}^{(r)}$, the value of $a_ir$ can be easily group into a small number of intervals, where these intervals has a large distinguishable gap between one another. Second, $K$-means clustering is fast, simple and easy to implement. With Silhouette score, we can automatically decide $K$ and obtain a good clustering result.

**Silhouette Clustering Criterion** The Silhouette score graphically measures how well a point lies within its cluster [23]. Assume that we have $K$ clusters $\{C_k\}, (k = 1, \cdots, K)$ and a similarity function $s(i, j)$ for data points $i$ and $j$. The similarity of a node $i$ to a cluster $C_k$ is defined via average similarity

$$d(i, C_k) = \sum_{j \in C_k} s(i,j)/|C_k| \ .$$

Suppose that node $i$ is in cluster $C_k$, then the *Silhouette score* of $i$ is

$$S(i) = \frac{d(i, C_k) - \max_{l \neq k} d(i, C_l)}{\max_m d(i, C_m)} \ \in [-1, 1]$$

Here, $S(i)$ measures how well $i$ belongs to its own cluster, with positive values indicating good clustering.

Finally, the average Silhouette score over all data points, $\sum_{i=1}^{|V|} S(i)/|V|$, is a measure of the quality of the $K$ clusters. The Silhouette criterion is to choose $K$ with the highest average Silhouette score.

Suppose $K^{(r)}$ clusters are detected in the generative model $X^{(r)}$, with the $m$-th cluster denoted as $C_m^{(r)}$. In our model, similar nodes generating more edges are more relevant to meaningful clusters. The $K$-means clustering in our framework can produce a cluster $C_m^{(r)}$ where $\forall i, j \in C_m^{(r)}, s(i, j) \approx 0,$. This means that $C_m^{(r)}$ generates almost no edges. Clusters such as $C_m^{(r)}$ are not considered to be meaningful and will be filtered by a cluster ranking step described next.

### 3.3 Cluster Ranking

To identify meaningful clusters, we rank $C_m^{(r)}$ for $m = 1, \cdots, K^{(r)}, r = 1, \cdots, R$, by their *average similarity ordering (SO)* score, which we define as:

$$SO_m^{(r)} = \frac{\sum_{i,j \in C_m^{(r)}} a_{ir} a_{jr}}{|C_m^{(r)}|^2} \int_0^T \lambda^{(r)}(z)dz, \qquad (5)$$

$$\approx \frac{\sum_{i,j \in C_m^{(r)}} a_{ir} a_{jr}}{|C_m^{(r)}|^2} (\sum_{t=1}^T \hat{\lambda}_{t,r}), \qquad (6)$$

where $\hat{\lambda}_{t,r}$ is the samples of edge-generating rate obtained from $\hat{X}^{(r)}$, as is introduced in Section 3.1. SO score represents, on average, the number of edges generated from a node-pair in a cluster over time. We generate a ranked list of clusters with SO score in decreasing order. Clusters at the top of the list are considered to be significant and provide useful information for analysis. Moreover, the elbow rule [13] can be applied to the SO score to filter out insignificant clusters: we choose the rank at which there is a significant drop in the SO score and remove clusters below it.

## 3.4 Lifetime Detection

This subsection describes the process: (1) constructing edge-generating rate $\lambda^{(r)}(t)$ from $T$ samples $\vec{\lambda}_r$ using time series segmentation for lifetime detection and (2) calculating the lifetime for a cluster.

### 3.4.1 Constructing $\lambda^{(r)}(t)$

We use $\vec{\lambda}_r$ from PARAFAC to construct the edge-generating rate $\lambda^{(r)}(t)$ for generative model $X^{(r)}$ for $r = 1, \ldots, R$.

We propose an approach based on time series segmentation with linear regression to construct $\lambda^{(r)}(t)$: given an integer $D$ and a time series $\vec{\lambda}_r = [\lambda_{1r}, \cdots, \lambda_{Tr}]$, let $\pi_D$ be a segmentation consisting of $D$ time segments $Q_d$ on time interval $[1, T]$. We define a piecewise linear model:

$$f_D(t) = \begin{cases} b_1 t + c_1, & t \in Q_1 \\ b_2 t + c_2, & t \in Q_2 \\ \vdots & \vdots \\ b_D t + c_D, & t \in Q_D \end{cases} \tag{7}$$

The objective is to find optimal $M_D^* = \{\pi_D, f_D^*\}$ that minimizes the squared error

$$e(\vec{\lambda}, M_D) = \sum_{j=1}^{T} (\lambda_{jr} - f_D(j))^2 \tag{8}$$

Then $\lambda^{(r)}(t)$ is represented by $f_D^*(t)$ in the solution $M_D^*$.

We need to decide the number of segments $D$. Note that $C(M_D^*)$ is sensitive to $D$: in the extreme case, the cost is 0 if $D = T - 1$ where each time step is a segment; at the other extreme when $D = 1$, it corresponds to the error of a linear regression on the entire time series. The most common methods use bottom up algorithms with a stopping criterion to set $D$. A typical algorithm begins with $T - 1$ segments and obtains a solution $M_{T-1}$. At each iteration, it merges two neighboring segments so as to minimize the increase in cost (denoted as $\Delta C(D) = e(\vec{\lambda}, M_{D-1}^*) - e(\vec{\lambda}, M_D^*)$).

Let $E_d$ be the squared error of using $f_D(t)$ to approximate $\vec{\lambda}$ in time segment $Q_d$, for $d = 1, \ldots, D$. One stopping criterion [12] defines a maximum error $E_{\max}$ such that $E_d \leq E_{\max}$, for $d = 1, \cdots, D$. However, it is difficult to choose the of threshold $E_{\max}$ for a given time series.

We assume $\vec{\lambda}_r$ contains noise, $\delta$, with normal distribution $N(\bar{\delta}, \sigma)$. The error $E_d$ is caused by $\delta$. We use $\bar{\delta} + 3\sigma$ as threshold for $E_{\max}$ because 99.7% of $\delta$'s values lie within three standard deviations of $\bar{\delta}$. The next step is to estimate the mean and variance of $\delta$. We apply a Sliding Window Filter to $\vec{\lambda}_r$ to compute $\hat{\lambda}_r(t)$, then obtain noise samples $\delta_j = \hat{\lambda}_r(j) - \lambda_{jr}$ for $j = 1, \cdots, T$, and calculate the mean $\bar{\delta} = \frac{1}{T} \sum_{j=1}^{T} \delta_j$ and standard error $\hat{\sigma} = \sqrt{\frac{1}{T-1} \sum_{j=1}^{T} (\delta_j - \bar{\delta})^2}$.

Finally, the *time series segmentation problem* can be formalized as minimize D:

$$\min D \tag{9}$$
$$s.t. \quad E_d \leq (\bar{\delta} + 3\hat{\sigma})^2 \quad \text{for } d = 1, \ldots, D$$

In Algorithm 1, the function LinearRegressionError($Y[i:j]$) calculates the linear regression $f_D(x)$ for a segment $Q_d = Y[i:j]$ and returns the error. The edge-generating rate $\lambda^{(r)}(t)$ can be constructed as $f_D(x)$ from the Algorithm 1 for $d = 1, \ldots, D$.

**Algorithm 1:** Time Mode Segmentation Algorithm

**Data**: Time Series $Y = [y_1, \cdots, y_T]$
**Result**: Segmentation $Q = [Q(1), \cdots, Q(D)]$, $f_D(x)$
$Q = [[1], [2], \cdots, [T]]$; $\hat{Y} = \text{SlideWindowFilter}(Y)$;
$\Delta = Y - \hat{Y}$; $\bar{\delta} = \text{mean}(\Delta)$; $\hat{\sigma} = \text{standardError}(\Delta)$;
**for** $i = 1 : T - 1$ **do**
    merge_cost(i) = LinearRegressionError(Y[i:i+1]);
**while** $min(merge\_cost) \leq (\bar{\delta} + 3\hat{\sigma})^2$ **do**
    i = indexOf(min(merger_cost));
    Q(i) = merge(Q(i),Q(i+1));
    delete(Q(i+1));
    merge_cost(i) = LinearRegressionError(merge(Q(i), Q(i+1)));
    merge_cost(i-1) = LinearRegressionError(merge(Q(i-1), Q(i)));

### 3.4.2 Lifetime Threshold

To detect lifetime of a cluster, we first define "*Average Edge-generating Rate for cluster $C_m^{(r)}$ at $t$*":

$$\lambda_m^{(r)}(t) = \frac{1}{|C_m^{(r)}|^2} \sum_{i,j \in C_m^{(r)}} a_{ir} a_{jr} \lambda^{(r)}(t) \tag{10}$$

Similarly, we define the "*Average Edge-generating Rate for Network $G$ at $t$*" as

$$\lambda(t) = \frac{1}{|V|^2} \sum_{r=1}^{R} \sum_{i,j \in V} a_{ir} a_{jr} \lambda^{(r)}(t) \tag{11}$$

Our model decides that a cluster $C_m^{(r)}$ appears at time $t$ if a node pair in $C_m^{(r)}$, on average, is expected to generate more edges than a randomly selected node pair from the network. The average edge-generating rate of node pairs in the network, $\lambda(t)$ (Eq (11)), serves as a threshold to decide the lifetime of a cluster. We define the lifetime of $C_m^{(r)}$ as:

$$L_m^{(r)} = \{t | \lambda_m^{(r)}(t) > \lambda(t)\} \tag{12}$$

where $\lambda_m^{(r)}(t)$ is the edge-generating rate of $C_m^{(r)}$ defined in Eq (10), $\lambda(t)$ can be easily obtained from the result of applying Sliding Window Filtering on the sequence $[\frac{\sum_{i,j=1}^{|V|} \mathcal{X}_{ijt}}{|V^2|}]$ for $t = 1, \cdots, T$.

## 4 Experiments and Evaluations

In this section, we evaluate our temporal clustering (TC) method. We first introduce the baseline methods: evolutionary clustering (EC) and PARAFAC with binary classification (BC). Then we describe the metrics for evaluating the methods. In Section 4.3, we use synthetic data to demonstrate the advantage of our model in obtaining ground truth clusters and determine their lifetimes by comparing its performance to EC [6] and BC [10]. Finally, we apply our method to the Enron data for community detection and lifetime detection.

### 4.1 Baseline Methods

Let $\mathcal{X}_t = \mathcal{X}_{(:,:,t)}$ be a similarity matrix for graph $G_t$ of $N$ nodes. EC performs clustering on the similarity matrix $\mathcal{R}_t = (1 - \beta)\mathcal{X}_t + \beta \mathcal{X}_{t-1}$ at time $t$ to obtain a clustering $\mathcal{C}_t = [\mathcal{C}_{1,t}, \ldots, \mathcal{C}_{N,t}]$, where $\beta \in [0, 1]$ and $\mathcal{C}_{i,t}$ is the index of the cluster that node $i$ belongs to at time $t$. Since $\mathcal{R}_t$ differs from the current similarity matrix $\mathcal{X}_t$, $\mathcal{C}_t$ is not necessarily the best clustering for $\mathcal{X}_t$, a snapshot quality function $sq(\mathcal{C}_t, \mathcal{X}_t)$ is used to evaluate the clustering result. EC includes a historical cost function $hc(\mathcal{C}_t, \mathcal{X}_{t-1})$ to calculate the difference between current and previous clustering results. The objective is to minimize

$$\sum_{t=1}^{T} sq(\mathcal{C}_t, \mathcal{X}_t) - c \sum_{t=2}^{T} hc(\mathcal{C}_t, \mathcal{X}_{t-1})$$

where $c$ is a scalar.

PARAFAC with BC considers each component $X^{(r)}$ as a community. It chooses a threshold for binary classification: For a node $i$, if $a_{ir}$ from the loading vector $A_r$ is larger than the threshold, then $i$ belongs to that community.

We run the EC algorithm with the program provided in public by Xu et. al [28], and use the tensor toolbox [4] for PARAFAC decompositon on a cluster of 32 Intel Xeon E5-2670@2.60GHz CPUs with 256Gb RAM. Additional code can be obtained through supplemental material on arXive (https://arxiv.org/abs/1605.08074).

## 4.2 Performance Metrics

We generate synthetic data with ground truth clusters to compare our method to the baseline methods. Since clusters identified by the methods mentioned above are different, we use clusters recall, cluster member recall, and precision-recall (PR) curves for evaluation instead of ROC curve.

When self-loop is allowed, we define the density of the $t$-th network snapshot as $\frac{2\sum_{i,j} \mathbf{1}(\mathcal{X}_{ijt} > 0)}{|V|^2}$, where $\mathbf{1}(\mathcal{D}) = 1$ if $\mathcal{D}$ is true, otherwise $\mathbf{1}(\mathcal{D}) = 0$.

### 4.2.1 Mapping Clusters to Ground Truth Clusters

We need to map the retrieved clusters to ground truth clusters before evaluation. A distance function between two clusters is required for the mapping. For evolving clulsters, the distance function should consider the dynamics of clusters.

Suppose that a method retrieves a ranked list of $K$ clusters $\{\hat{C}_i\}_{i=1}^{K}$, where $\hat{C}_i$ has higher rank than $\hat{C}_j$ if $i < j$. $\{C_n^*\}_{n=1}^{N}$ is the set of ground truth clusters. We use $x_{ijt}^*$ to denote the number of edges generated between $i, j \in C_n^*$ at $t$. $x_{ijt}^* = 0$ if $i, j$ belong to different ground truth clusters. We define the distance function between $\hat{C}_i$ and $C_n^*$ as the approximation error:

$$E(\hat{C}_i, C_n^*) = \frac{\sum_{t=1}^{T} \sum_{m,o \in \hat{C}_k \cup C_n^*} (b_{mr} b_{or} \hat{\lambda}^{(r)}(t) - x_{mot}^*)^2}{\sum_{t=1}^{T} \sum_{j,p \in C_n^*} (x_{jpt}^*)^2}$$

where $b_{mi} = a_{mi}$ if node $m \in \hat{C}_i$ and $b_{mi} = 0$ otherwise. Note that $E(\hat{C}_k, C_n^*) = 0$ if $\hat{C}_i = C_n^*$ and $E(\hat{C}_i, C_n^*) = 1$ if $\hat{C}_i$ is the empty set. We map $\hat{C}_i$ to the empty set $\phi$ if $E(\hat{C}_i, C_n^*) \geq 1$ for any $C_n^*$. The mapping is performed via the following steps:

(1) Begin with $i = 1$ and map $\hat{C}_i$ to $C_n^* = \arg\min_{C_j^*} E(\hat{C}_i, C_j^*)$ for $j = 1, \ldots, N$.

(2) Increase $i$ by 1 and repeat the previous step. Note that two clusters, $\hat{C}_i$ and $\hat{C}_j$, may be mapped to the same ground truth cluster $C_n^*$, for $i < j$. We allow these two mappings only if $\hat{C}_i$ and $\hat{C}_j$ are generated from the same generative model $X^{(r)}$. In this case, $\hat{C}_i$ and $\hat{C}_j$ are likely to be sub-clusters of $C_n^*$. Otherwise, we consider $E(\hat{C}_j, C_n^*) = 1$ and choose another $C_{n'}^*$ for $C_j$.

### 4.2.2 Precision, Recall, F1 Measure and PR Curve

Precision, recall and F1 Measure (or F1 score) are common metrics to evaluate clustering when ground truth knowledge is available. We apply these metrics to evaluate the performance of EC, BC and TC.

Suppose a detected cluster $\hat{C}_k$ is mapped to ground truth cluster $C_n^*$, the precision for member in cluster is $\mathcal{P} = \frac{|\hat{C}_k \cap C_n^*|}{|\hat{C}_k|}$, the member recall is $\mathcal{R} = \frac{|\hat{C}_k \cap C_n^*|}{|C_n^*|}$, and the F1 score is defined as the harmonic mean of $\mathcal{R}$ and $\mathcal{P}$: $\mathcal{F}_1 = \frac{2\mathcal{P}\mathcal{R}}{\mathcal{P}+\mathcal{R}}$.

If a method detects $P$ out of $N$ ground truth clusters, the cluster recall is $\frac{P}{N}$. Suppose $M$ clusters identified as meaningful by the method, the cluster precision is $\frac{P}{M}$. The F1 score for clusters is the harmonic mean of cluster recall and cluster precision.

Since the number of cluster identified by our method may be different from the baseline methods, we also use precision-recall (PR) curve for comparison. A PR curve is usually applied when the compared methods have similar recall scores. A PR curve requires a clustering method to return a rank list of clusters. Suppose the rank list has $K$ clusters, $\{\hat{C}_i\}_{i=1}^k$ are top $k$ clusters in the list and are mapped to $k$ out of $N$ ground truth clusters $\{C_{n_i}^*\}_{i=1}^k, 1 \leq n_i \leq N$. For $k = 1, \cdots, K$, we define precision of the top $k$ clusters as $\mathcal{P}_k = \frac{\sum_{i=1}^k |\hat{C}_i \cap C_{n_i}^*|}{\sum_{i=1}^k |\hat{C}_i|}$ and recall as $\mathcal{R}_k = \frac{\sum_{i=1}^k |\hat{C}_i \cap C_{n_i}^*|}{\sum_{j=1}^N |C_j^*|}$. We generate a PR curve by plotting $(\mathcal{P}_k, \mathcal{R}_k)$ in the PR space. $\mathcal{R}_K$ is defined as the *cluster member recall for a dynamic network*.

To generate a ranked list from BC, let $A_r = [a_{1r}, \ldots, a_{|V|r}]$ denote the loading vector in the $r$th component $X^{(r)}$. BC classifies a node $i$ as a member in a community of interest, $\hat{C}_1^{(r)}$, if $a_{ir}$ is larger than a threshold. Otherwise $i$ is put in another community $\hat{C}_2^{(r)}$ that can be ignored. Components $X^{(r)}$ are placed in descending order of their Frobenius norm, for $r = 1, \ldots, R$. $\hat{C}_1^{(r)}$ is believed to have stronger structures than $\hat{C}_1^{(r')}$ if $r < r'$ [10]. A ranked list from BC can be generated as: $[\hat{C}_1^{(1)}, \ldots, \hat{C}_1^{(R)}, \hat{C}_2^{(1)}, \ldots, \hat{C}_2^{(R)}]$

### 4.2.3 Metric for LifeTime Detection

We use $L_m^{(r)}$ in Eq (12) to classify a time step $t$ as being in the lifetime of cluster $C_m^{(r)}$. If an algorithm detects a cluster, we use F1 score of lifetime steps for evaluation.

## 4.3 Evaluation with Data

The synthetic data includes randomly generated dynamic networks with multiple clusters and a mobility network generated from Lakehurst mobility trace data set from [7]. The real-world data is a dynamic network generated from Enron email dataset[22].

### 4.3.1 Synthetic networks

We generate 5040 synthetic dynamic networks to evaluate our temporal clustering framework and compare its performance to EC and BC on community detection and lifetimes detection at different time granularities.

Network sizes range from 100 to 500, with a number of snapshots $T \in [1000, 4000]$. In a network, clusters may have common nodes and their sizes vary from 8 to 80. The number of clusters in a network is sampled uniformly from $[10, 40]$.

For each cluster, we split a time interval $[1, T]$ into multiple segments and choose a fraction of those segments as the lifetime of the cluster. During each lifetime segment, we set the edge-generating rate as either a constant or a linear function, with its value varying from 0.0015 to 1. We set the lifetime of a cluster to be at least $0.3T$ time steps such that there are enough edges generated from the cluster for clustering. We also set up periodic lifetimes for some clusters. The periods range from 20 time steps to $\frac{T}{2}$.

Finally, we add noise to each network: each node pair in a network generates an edge with rate $e$ during a period of one time step, where $e$ is sampled uniformly from $[0, 0.01]$.

We also examine the effect of time granularity on the performance of different methods. In particular, we consider datasets where $w$ continuous snapshots are aggregated together. In such case, an original dataset consisting of T snapshots is transformed into one consisting of $T/w$ snapshots. At the smallest granularity, let $\mathcal{X}_{ijt}$ be the weight or number of edge between node $i, j$ at time step $t$. At time granularity $w$, the weight of edge between node $i, j$ is defined as:

$$\mathcal{X}_{ijt'}(w) = \sum_{t=(t'-1)w}^{t'w} \mathcal{X}_{ijt}$$

We apply EC [28], BC and our method to each synthetic network.

**Evaluation** We first use F1 score to evaluate the performance of methods that retrieve communities in networks given different granularity $w$. We set $w = 2^q$ for $q = 0, \ldots, 7$. We define the density of a cluster $C_k$ at $t$ as $d_k(t) = \frac{2\sum_{i,j \in C_k}(a_{ir}a_{jr}\lambda^{(r)}(t))}{|C_k|^2}$, and use $d_k$ to denote the average density through out lifetime. Note that the edge-generating rate $\lambda^{(r)}(t)$ affects the density of a cluster greatly.

Average cluster densities in our synthetic data vary from 0.0008 to 0.9341. We split networks into 10 groups according to their average cluster density. Networks in the $i$-th group have average cluster density within $[(i-1) \times 0.1, i \times 0.1]$. Let $K$ denote the number of ground truth clusters in a synthetic network, we set the rank $R = 0.3K, 0.4K, \ldots, K$ for BC and our temporal clustering method. We illustrate their performance at rank $R_1 = K, R_2 = 0.8K$ and $R_3 = 0.6K$.

Figure 1 shows the change in F1 score (F1 measure) of clusters as a function of time granularity. In Figure 1(a), the three methods identify clusters in networks whose average cluster densities falls in interval $[0.3, 0.4]$. The F1 scores for both BC and our temporal lustering method (TC) change little with the increase of granularity. EC retrieves no ground truth clusters when the granularity is small, but begin to identify clusters as granularity increases. BC and TC perform better when the rank $R$ is set to $K$. As $R$ decreases to $0.8K$ and $0.6K$, the F1 scores of both methods drop significantly. BC is more sensitive to change in $R$. Figure 1(b) illustrates the result on networks whose average cluster density ranges from 0.7 to 0.8. Compared to Figure 1(a), the performances of BC and TC are similar, but the F1 score for EC improves as cluster densities increase. Figure 1 suggests that EC is sensitive to the change in granularity and the cluster density when detecting communities in dynamic network, while both BC and TC are more robust than EC. TC has similar performance as BC when $R$ is chosen properly, but performs better in identifying ground truth clusters than BC when $R$ is improperly small.

We next evaluate the performance on retrieving cluster members. We are interested in the effect of rank $R$, time granularity, and average cluster density in lifetime on the performance of the methods.

Figure 2 illustrates the F1 scores of identification of ground truth cluster members by EC, BC and TC. In Figure 2(a), we collect a set of detected clusters $\{C_k\}$ with cluster density $d_k \in [0.3, 0.4]$ across all the synthetic networks and calculate the F1 score of cluster members. F1 score for EC is larger than zero only when granularity is above 32. TC and BC both have F1 scores larger than that of EC. For each $R$ value (from $K, 0.8K$ to $0.6K$) TC has better F1 score than BC. Figure 2(b) shows the results for clusters with densities within $[0.9, 1]$, the high cluster density suggests that almost every node in the cluster is connectd to one another at each network snapshot. The performance of EC improves subsequently over that in Figure 2(a). Since we allow clusters to share common nodes in the network, EC only retrieves disjoint clusters, its F1 score is lower than that of TC and BC. On the other hand, TC and BC exhibit similar performance on clusters with different densities. TC does better than BC, especially as $R$ decreases.

**Lifetime Detection** We now focus on the quality of cluster lifetime detection provided by EC, BC and TC. Figure 3 shows the F1 score of lifetime detection for clusters with densities within $[0.7, 0.8]$ and $[0.9, 1]$. Lifetime detection by TC and BC is sensitive to change in time granularity. The reason is that information for lifetime detection is lost when network snapshots are aggregated at coarse granularities. The F1 scores for both TC and BC stop decrease as time granularity increases to a certain value. Lifetime detection by EC benefits from the increase in time granularity, but accuracy stops improving when granularity becomes large.

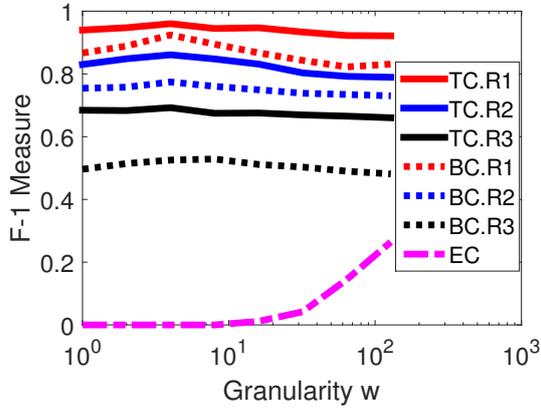
(a) Networks with Average Cluster Density in $(0.3, 0.4)$

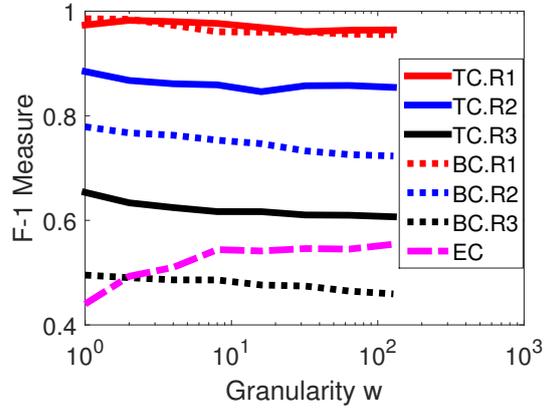
(b) Networks with Average Cluster Density in $(0.7, 0.8)$

Figure 1: F1 Measure for Cluster Identification. Let $K$ be the number of ground truth clusters in a network, BC and temporal clustering (TC) apply PARAFAC with rank $R_1 = K$, $R_2 = 0.8K$, and $R_3 = 0.6K$

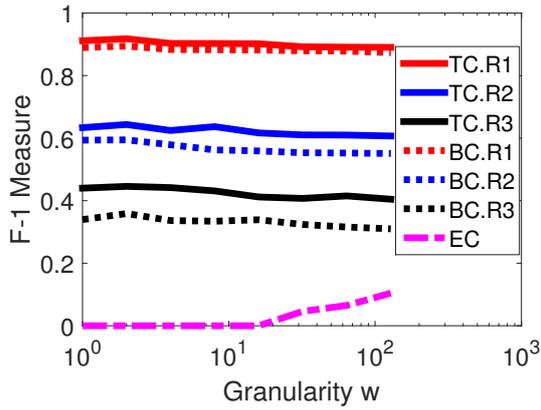
(a) Clusters with Average Density in $(0.7, 0.8)$

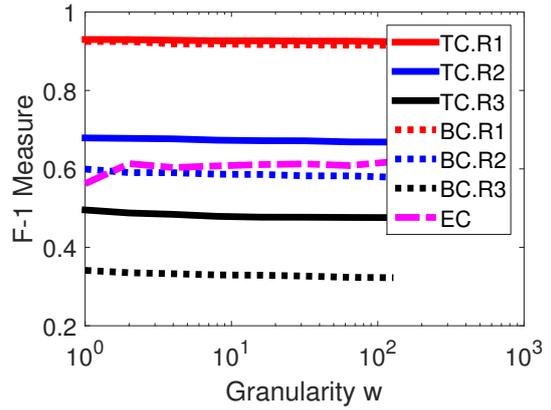
(b) Clusters with Average Density in $(0.9, 1)$

Figure 2: F1 Measure for Cluster Member Identification. BC and TC apply PARAFAC with rank $R_1 = K$, $R_2 = 0.8K$, and $R_3 = 0.6K$

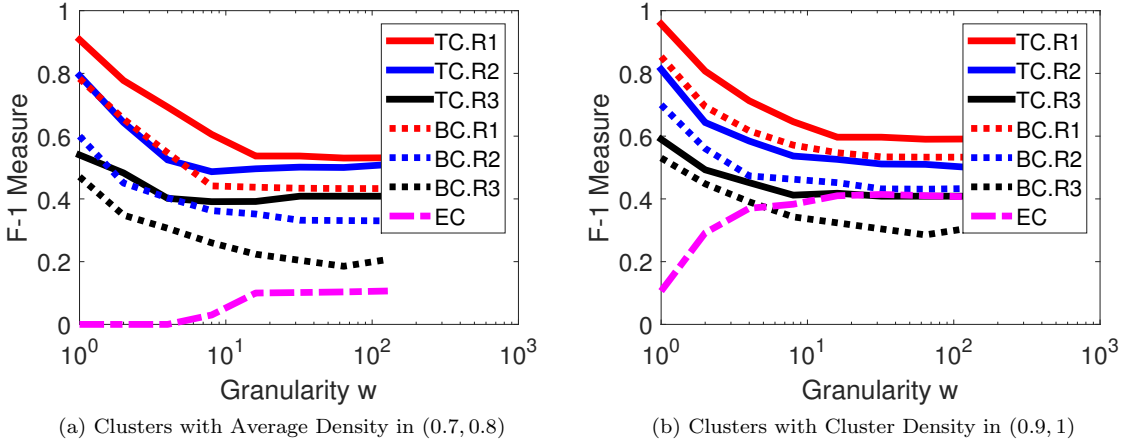

(a) Clusters with Average Density in $(0.7, 0.8)$

(b) Clusters with Cluster Density in $(0.9, 1)$

Figure 3: F1 Measure for Lifetime Detection. BC and TC apply PARAFAC with rank $R_1 = K$, $R_2 = 0.8K$, and $R_3 = 0.6K$

We are interested in the effect of granularity $w$ on infering periodic lifetimes. Figure 4 shows the detection of periodicity in lifetime for two ground truth clusters $C_1^*$ and $C_2^*$ taken from the same dynamic network, as well as their edge-generating rates when $w = 1$ and $w = 128$. Figure 4(a) shows lifetime detection for $C_1^*$ by our method. The ground truth edge-generating rate is denoted as $\lambda_1^*(1, t)$. Its sample at time step $t$ from TC at granularity $w$ is $\lambda_{1,t}(w)$, for $w = 1, 128$. At granularity $w$, we reconstruct the edge-generating rate as $\lambda_1(w, t)$, using samples $\lambda_{1,t}(w)$. $\lambda(t)$ in Eq(11) is denoted as $\lambda(w, t)$. The lifetime of $C_1^*$, denoted as $L_1(w)$, is obtained with Eq(12). Note that the period of the lifetime of $C_1^*$ is small and we can maintain the periodicity at $w = 1$. At $w = 128$, however, lifetime $L_1(128)$ is not accurate and the periodicity is lost. Figure 4(b) illustrates the lifetime of cluster $C_2^*$. The period is larger than that of $C_1^*$. The periodicity at both granularities $w = 1$ and $128$ can be maintained.

### 4.3.2 Lakehurst Data

The Lakehurst data set from the US Army Research Lab contains a three hour (10800 seconds) trace of 70 vehicles (ground and airborne) [7]. 64 vehicles are split into 9 platoons, moving from one checkpoint to another. At the end, all vehicles reach the same destination. There are two paths from the start location to the destination and each path has five checkpoints. A platoon chooses one path and stops for a while at each checkpoint until it reaches the destination. The other six vehicles move separately but sometimes have intersecting paths with one of those platoons.

The data set contains the location of each vehicle every second. Vehicles in a platoon move together and two vehicles in the same platoon are within 150 meters for 99% of the time. We create a dynamic graph by adding an edge between two vehicles within 150 meters of each other at each time step. In each snapshot of the graph, nodes in the same platoon tend to have an edge. The average density of the adjacency matrix of a snapshot of is 0.235 and the maximum is 0.688. The network structure changes slowly, sometimes it could remain unchanged for 3 minutes. We construct a $70 \times 70 \times 10800$ tensor. Some platoons meet one another and periodically form a larger cluster in a period of time. There are 10 such clusters. As a result, there are 19 ground truth clusters appearing in the dynamic network. The size of clusters ranges from 5 to 25.

Since platoons have intersecting path and sometimes forms larger clusters only for seconds, we are interested in the number of ground truth clusters that a method obtains. We use cluster recall and cluster member recall to evaluate the performance of EC, BC and TC. We applied PARAFAC

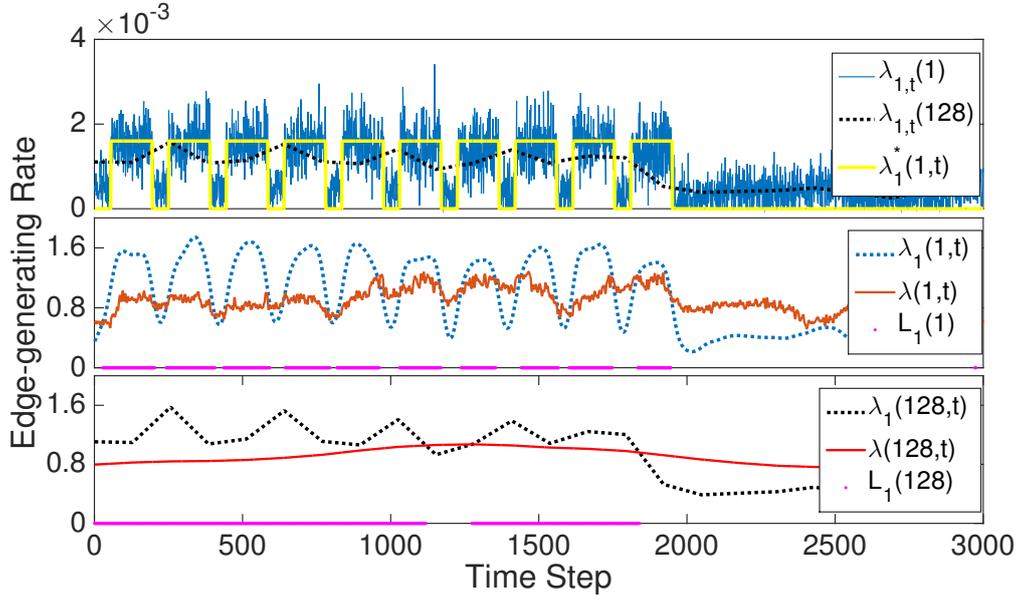

(a) Lifetime Detection of $C_1^*$ at $w = 1, 128$

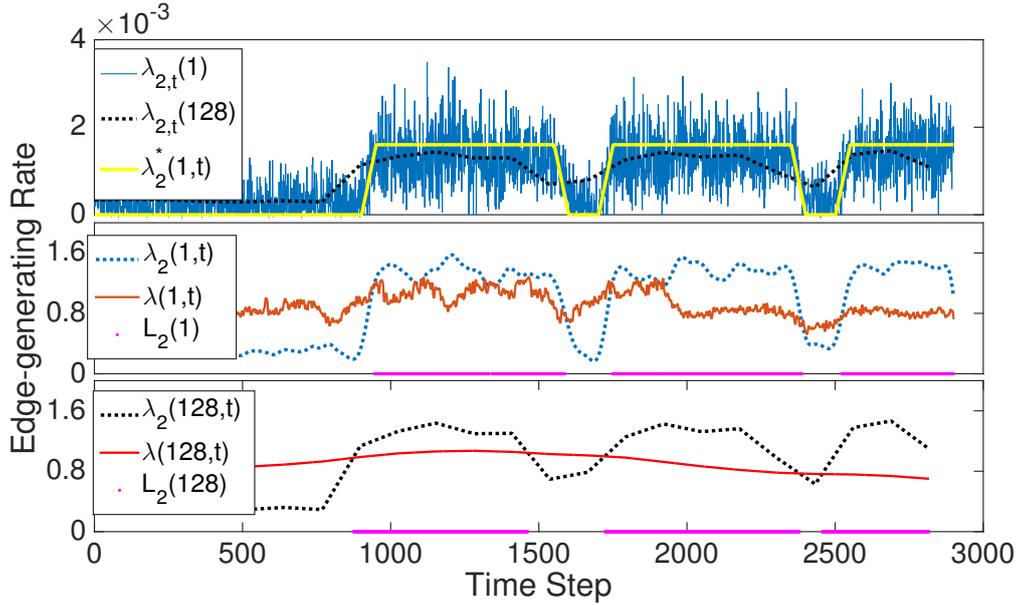

(b) Lifetime Detection of $C_2^*$ at $w = 1, 128$

Figure 4: Periodic Lifetime Detection for Two Clusters $C_1^*$ and $C_2^*$ at Granularity $w = 1, 128$. (a) Periodic lifetime with short period loses periodicsity at large time granularity $w = 128$. (b) Lifetime with long period maintains its periodicity at $w = 128$.

decomposition with $R = 10, 15, 20, 25$ and $30$. The recall of ground truth clusters and nodes are illustrated in Table 1.

The snapshots in this data set is dense. EC detects clusters and cluster members even at the smallest time granularity. However, it only retrieves the nine ground truth platoons and fails to obtain the larger clusters formed by multiple platoons.

Table 1: Clusters Recall and Member Recall for Lakehurst Data

|  | Temporal Clustering | | | | | PARAFAC w/ BC | | | | | EC |
| --- | --- | --- | --- | --- | --- | --- | --- | --- | --- | --- | --- |
| $R$ | 10 | 15 | 20 | 25 | 30 | 10 | 15 | 20 | 25 | 30 | N/A |
| Cluster Recall | 0.368 | 0.421 | 0.579 | 0.895 | 1.0 | 0.319 | 0.421 | 0.579 | 0.895 | 1.0 | 0.474 |
| Member Recall | 0.430 | 0.541 | 0.841 | 0.875 | 0.904 | 0.403 | 0.532 | 0.841 | 0.862 | 0.904 | 0.275 |

Figure 5 illustrates the PR curves generated from the results of BC and TC with SO ranking with PARAFAC decomposition rank $R = 15, 30$. Both sub-figures show that SO ranking in TC has better precision than BC when the recall is less than 0.5, suggesting the ordered list by SO ranking is more accurate than BC. The reason is that in a component with larger norm, some nodes $i$ do not belong to any cluster, but have positive values $a_{ir}$ larger than the threshold in BC. BC simply considers that a component only contains one meaningful cluster and mistakenly classifies $i$ as a cluster member. Our method seperates nodes into multiple clusters and give low rankings to clusters with small SO scores. Moreover, some components from PARAFAC are too erroneous to be suitable for community detection, our method gives clusters from those components low SO scores.

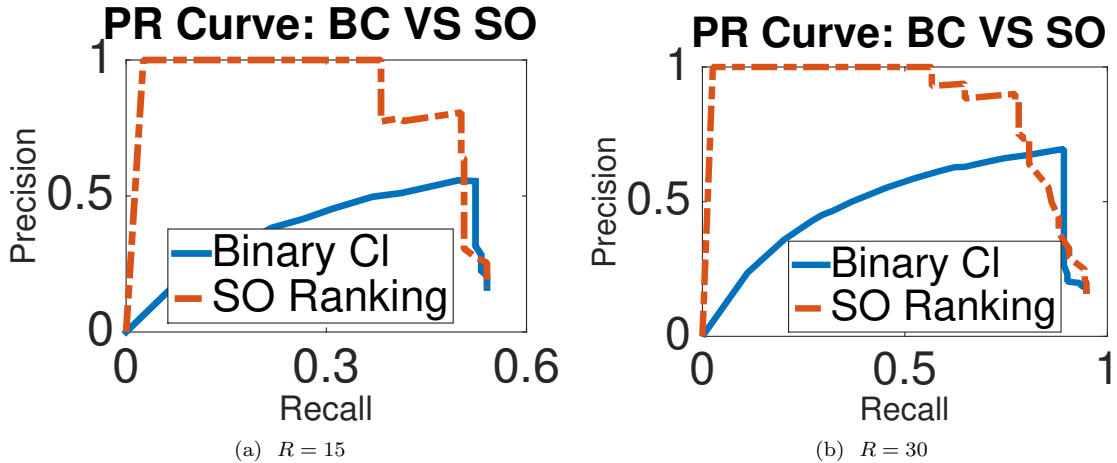

(a) $R = 15$

(b) $R = 30$

Figure 5: PR Curves for BC and SO Ranking Performance.

## 4.4 Application to Enron Email Data

We apply our method to a dynamic network constructed from the Enron email dataset by Priebe et al. [22]. The Enron dataset contains 184 unique email addresses and 125,409 messages dated from November 1998 to June 2002. We represents the date as a time varying undirected graph with time granularity of one second, resulting in a total of $113,733,036$ time steps. This produces a tensor that exceeds the capacity of our computational resource. Hence we construct a $184 \times 184 \times 31,592$ tensor with a time granularity of one hour $w = 3600$. We denote the tensor as $\mathcal{X}$, where $\mathcal{X}_{ijt}$ is the number of emails exchanged between person $i$ and person $j$ at the $t$-th hour. The average density of a matrix of a snapshot is $3.22 \times 10^{-5}$ and the maximum density is 0.0019. The edge-generating rate here corresponds to the email-exchanging rate.

We apply PARAFAC with rank $R = 60$ to $\mathcal{X}$ and generate a ranked list of 120 clusters using SO score. There is a large gap between the SO score of the 60-th cluster and 61-th cluster; hence we apply the elbow rule to decide that the first 60 clusters are worth investigating.

Since ground truth is not known for Enron dataset, we refer to results stated in [8, 27] to validate our results. We also check if our method provides new insights into behaviors that produced by email exchanges.

One interesting discovery from our method are differences of the email exchange behavior of different communities. Figure 6 shows the average weekly email exchange rate of two groups. The first community in our ranked list consists of two CEOs, three presidents and an employee. These people exchanges emails even on weekends. According to Diesner et. al [8], people in this community are "key players" with respect to "centrality measure". The employee has node ID 64 in the dataset and turns out to be Jeff Dasovich[22], who is known to exchange a lot of emails and has the highest eigenvector Centrality. On the other hand, the fifth community detected in our model consists of a president and other employees. They only communicate during working hours on weekdays.

We discover that some people belong to different communities. For example, Richard Shapiro, the vice president of regulatory affair, and Jeff Dasovich appear together in different groups. All identified communities exhibit weekly periodic lifetime and their edge-generating rates have similar patterns every weekday. Some communities have lifetime across three years period (e.g., a community including presidents (Kevin Presto, richard Shapiro, Shelley Corman) and several employees (Jeff Dasovich, Kay Mann, Susan Scott)), while some last for less than a year (e.g., a community including a manager(Philip Allen) and employees(Daron Giron, phillip Love, and Eric Bass)).

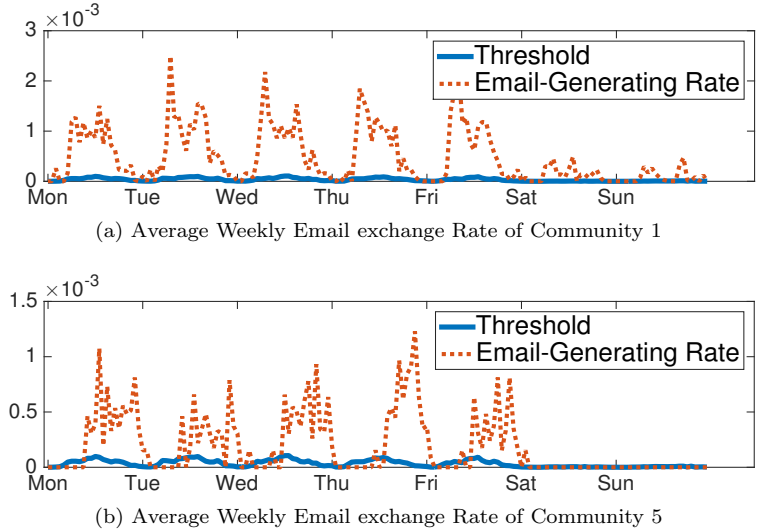

Figure 6: Different Email Exchanging Behavior between Identified Clusters.(a) Email exchange rate of a community consisting of two CEOs ( David Delainey and John Lavorato), three vice presidents(James Steffs, Richard Sanders nd Richard Shapiro) and an Employee (Jeff Dasovich) (b) Email exchange rate of a community consisting of a president (Louise Kitchen), director(Jonathon Mckay), Employees (Cara, Kate) and traders (Diana, Ryan). The first community exchanges emails even on weekends while the second only exchange emails on weekdays. Time steps belong to a community lifetime when email-generatig rate is larger than theshold.

# 5 Conclusion

We proposed a temporal clustering method based on a network generative model to detect clusters and obtain their temporal information in a dynamic network. PARAFAC decomposition is applied to learn the model and $K$-mean clustering with Silhouette Criterion is used to obtain communities. We propose a time series segmentation algorithm to analyze change in clusters and provide a way to estimate cluster lifetimes. Experiments show that our method has advantages over evolutionary clustering and other PARAFAC decomposition methods such as BC when snapshots are too sparse to provide cluster structure or lifetime information.

Several insights from our experiments and analysis could be helpful to the research on community detection of dynamic networks:

(1) A snapshot-based clustering model such as evolutionary clustering uses local information from a number of network snapshots. From our experiments with synthetic data , we observed that it is effective if a network is sufficiently dense. Our model and other PARAFAC decomposition methods takes advantage of global information of all snapshots and usually obtain better clustering result.

(2) Deciding on the number of communities correctly in a dynamic network is challenging and important to the results. Our experiments suggest that if the rank of PARAFAC affects greatly the performance of BC and our methods. However, our temporal clustering method allows multiple clusters in one generative model and obtain more ground truth cluster and members given the same setting as BC.

(3) Processing data at coarse scale time granularities helps snapshot-based clustering, but is relatively less effective for models like PARAFAC decomposition that use global information. Temporal information such as periodicity in the lifetime of a cluster can be lost at large granularity.

(4) It is difficult to choose a proper granularity for lifetime detection, especially for methods such as EC, that cannot identify sparse clusters in small granularity. However, our method has good performance as long as a coarse granularity does not cause information loss.

(5) Experiment codes and more details on our model can be found through supplemental material on more detail can be found on github (https://github.com/submissionCode/KDD2017).

# A  Extended Model

## A.1  Modeling Networks without Self-loop

In Section 2, we introduce our model on networks that allow self-loop. In this section, we explain how to extend our model to networks without self-loops and how to learn the model.

Since a node $i$ does not generate a self-loop connecting itself, the probability of generating an edge in generative model $X^{(r)}$, denoted as $P^{(r)}(e_{ii})$, can be modeled as 0, where $e_{ii}$ represent an edge connecting node $i$ to itself. We first introduce a binary weight $W_{ij}$, where $W_{ij} = 1$ if $i \neq j$ and $W_{ij} = 0$ otherwise. Then we define

$$P^{(r)}(e_{ij}) = W_{ij} a_{ir} a_{jr} \lambda^{(r)}(t)$$

.

We still use a three mode $|V| \times |V| \times T$ tensor $\mathcal{X}$ to represent a network without self-loops, where $|V|$ is the number of nodes, $T$ is the number of snapshots and $\mathcal{X}_{ijt}$ represents the number (weight) of edge between node $i, j$ at time step $t$. Note that $\mathcal{X}_{iit} = 0$ for $i = 1, \ldots, |V|$. The objective, Eq (1), in Section 2 is:

$$\begin{aligned}
&\min \sum_{i,j \in V} \sum_t (\mathcal{X}_{ijt} - W_{ij} \sum_{r=1}^R a_{ir} a_{jr} \lambda^{(r)}(t))^2 \\
&\text{s.t} \quad 0 \leq a_{ir} \leq 1; \text{ for } i=1,\ldots,|V| \\
&\qquad\quad W_{ij} = 1, \text{ for } i \neq j \\
&\qquad\quad Wii = 0, \text{ for } i=1,\ldots,|V| \\
&\qquad\quad \lambda^{(r)}(t) \geq 0
\end{aligned} \qquad (13)$$

To learn this model, we apply CP weighted optimization algorithm[2] rewriting Eq (3) as

$$\min \sum_{i,j,t} W_{ijt}(\mathcal{X}_{ijt} - \sum_{r=1}^R a_{ir} a_{jr} \lambda_{tr})^2$$

.

## A.2 Networks at Different Granularities

Networks are analyzed at different time granularities due to researchers' objectives. The results at different granularities are sometime incomparable, for example, an edge-generative rate obtained at a fine granularity is smaller than that at a coarse granularity. To make our model applicable to different granularity, we extend the edge-generating rate as a function of both granularity and time.

We first define time granularity, $w$, as the number of continuous snapshots to aggregate so as to analyze a dynamic network. At the finest time granularity, let $\mathcal{X}_{ijt}$ represents the observed number of edges (or weight of the edge) between node $i,j$ at time step $t$. At granularity $w$, We define

$$\mathcal{X}_{ijt'}(w) = \sum_{t=(t'-1)w}^{t'w} \mathcal{X}_{ijt} \qquad (14)$$

where $t'$ is a time interval at granularity $w$ that includes time step $t \in [(t'-1)w, t'w]$.
The edge-generating rate for generative model $X^{(r)}$ is represented as

$$\lambda^{(r)}(w, t') = \int_{(t'-1)w}^{t'w} \lambda^{(r)}(x) dx \approx \sum_{t=(t'-1)w}^{t'w} \lambda^{(r)}(t) \qquad (15)$$

Given granularity $w$, the objective of our network generative model becomes

$$\begin{aligned}
&\min \sum_{i,j \in V} \sum_t (\mathcal{X}_{ijt}(w) - \sum_{r=1}^R a_{ir} a_{jr} \lambda^{(r)}(w,t))^2 \\
&\text{s.t} \quad 0 \leq a_{ir} \leq 1; \text{ for } i=1,\ldots,|V| \\
&\qquad\quad \lambda^{(r)}(w,t) \geq 0
\end{aligned} \qquad (16)$$

# B Experiment Results

## B.1 Additional Experiment Results for Sysnthetic Networks

In this section, we provide additional experiment results from TC and BC on the synthetic network in support of performance analysis in Sec 4. F1 scores for cluster detection, cluster member detection and lifetime detection from TC and BC are illustrated as a function of granularity, given different rank of PARAFAC decomposition.

Figure 7 shows the performance of cluster identification in networks by EC, BC and TC using F1 scores, given the average cluster densities. EC performance is sensitive to both the change in average

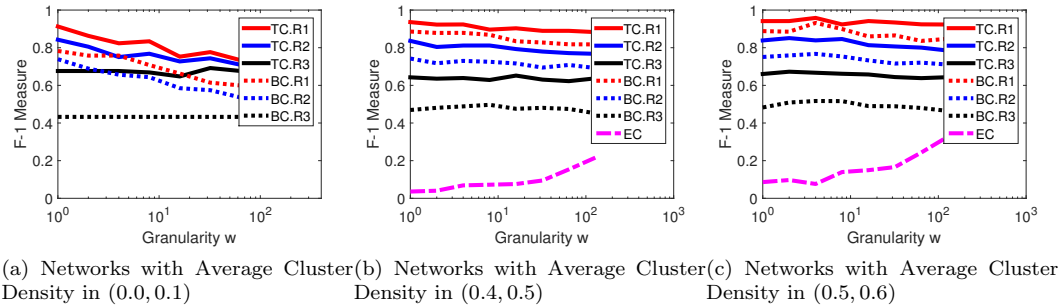

(a) Networks with Average Cluster Density in $(0.0, 0.1)$
(b) Networks with Average Cluster Density in $(0.4, 0.5)$
(c) Networks with Average Cluster Density in $(0.5, 0.6)$

Figure 7: F1 Measure for Cluster Identification in Synthetic Networks. Let $K$ be the number of ground truth clusters in a network, BC and temporal clustering (TC) apply PARAFAC with rank $R_1 = K$, $R_2 = 0.8K$, and $R_3 = 0.6K$. Clusters with low average density are difficult to detect. EC fails to identify clusters in any network with averge cluster density within $(0, 1)$. BC and TC can still detect clusters and are robust to the change of granularity. BC is sensitive to the change of rank of PARAFAC $R$. TC is less sensitive than BC and maintains much better performance than BC when $R$ is smaller than the best rank of PARAFAC

Table 2: Top Active Person in Enron Community 14

| Name | Position | Role |
| --- | --- | --- |
| Louise Kitchen | President | Enron Online |
| Mark Haedicke | Managing Director | Legal Department |
| Mark Taylor | Employee | |
| Stephanie Panus | Employee | |
| Susan Bailey | N/A | |
| Tana Jones | N/A | |

cluster density and granularity. Moreover, it has worse performance in networks whose clusters are allowed to overlap. BC and TC are both robust to change in average cluster density and granularity. When the rank of PARAFAC $R$ is smaller than the number of ground truth clusters in a network, both BC and TC have their F1 score decreased, but TC is less sensitive to the change in $R$ and has significant improvment on BC.

Figure 8 shows the performance of cluster member detection by BC and TC. PARAFAC are applied on all networks with rank $R = 0.3K, 0.5K, 0.7K, 0.9K, K, 1.5K$. We group clusters according to their densities and evaluates in each groups the F1 scores of BC and TC. TC has better performance when $R$ is improperly chosen.

Figure 9 shows the performance of lifetime detection of clusters by BC and TC, given different cluster densities. F1 scores for both BC and TC decreases with the increase in granularity. The cluster density has little effect on the F1 scores when it is above 0.3.

## B.2 Enron Data

In this subsection, we provide details of the first 15 Enron communities that are detected from our method.

From Figure (10,11,12), communities have different lifetime. For example Community 14 exist back in 1999 and last for more than 3 years until 2002. It contains Louise Kitchen (President), Mark Haediche(Director) and several employee (Table 2).

Community 15 forms near the bankruptcy in 2001 and only lasted for several months. It includes Mark haedicke and James Derrick who are related to legal department (Table 3).

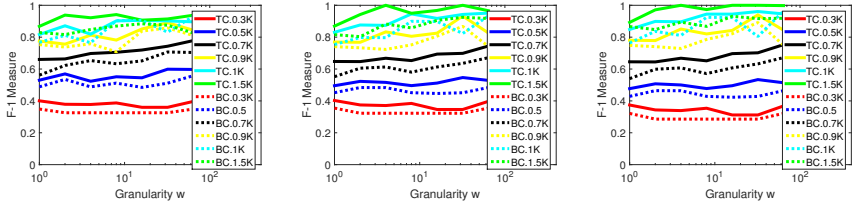

(a) Cluster with Density in $(0.2, 0.3)$
(b) Cluster with Density in $(0.3, 0.4)$
(c) Cluster with Density in $(0.4, 0.5)$

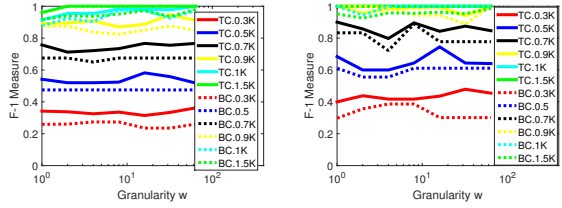

(d) Cluster with Density in $(0.5, 0.6)$
(e) Cluster with Density in $(0.6, 0.7)$

Figure 8: F1 Measure for Cluster Member Identification in Clusters. Let $K$ be the number of ground truth clusters in a network, BC and temporal clustering (TC) apply PARAFAC with rank $R = 0.3K, 0.5K, 0.7K, 0.9K, K, 1.5K$. BC and TC are robust to the change of granularity. BC is sensitive to the change of rank of PARAFAC $R$. TC is less sensitive than BC and maintains much better performance than BC when $R$ is smaller than the best rank of PARAFAC

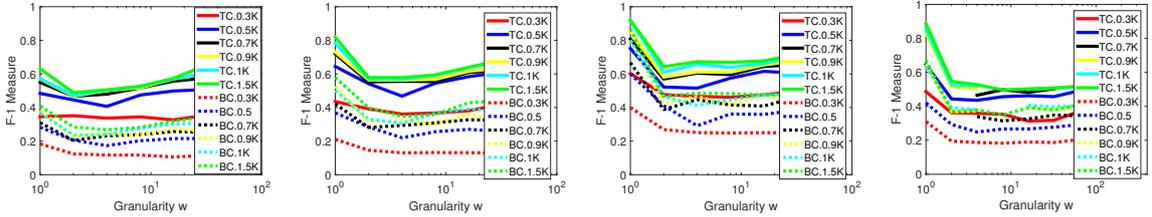

(a) Cluster with Density in $(0.0, 0.1)$
(b) Cluster with Density in $(0.1, 0.2)$
(c) Cluster with Density in $(0.2, 0.3)$
(d) Cluster with Density in $(0.3, 0.4)$

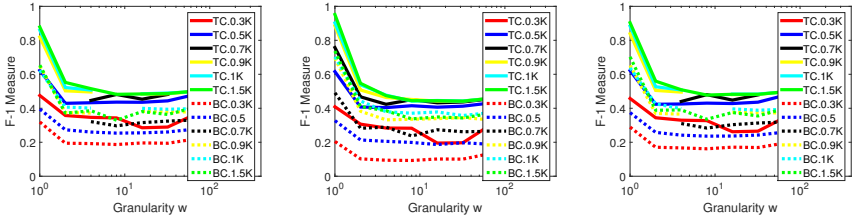

(e) Cluster with Density in $(0.4, 0.5)$
(f) Cluster with Density in $(0.5, 0.6)$
(g) Cluster with Density in $(0.6, 0.7)$

Figure 9: F1 Measure for Cluster Member Identification in Clusters. Let $K$ be the number of ground truth clusters in a network, BC and temporal clustering (TC) apply PARAFAC with rank $R = 0.3K, 0.5K, 0.7K, 0.9K, K, 1.5K$. BC and TC are robust to the change of granularity. BC is sensitive to the change of rank of PARAFAC $R$. TC is less sensitive than BC and maintains much better performance than BC when $R$ is smaller than the best rank of PARAFAC

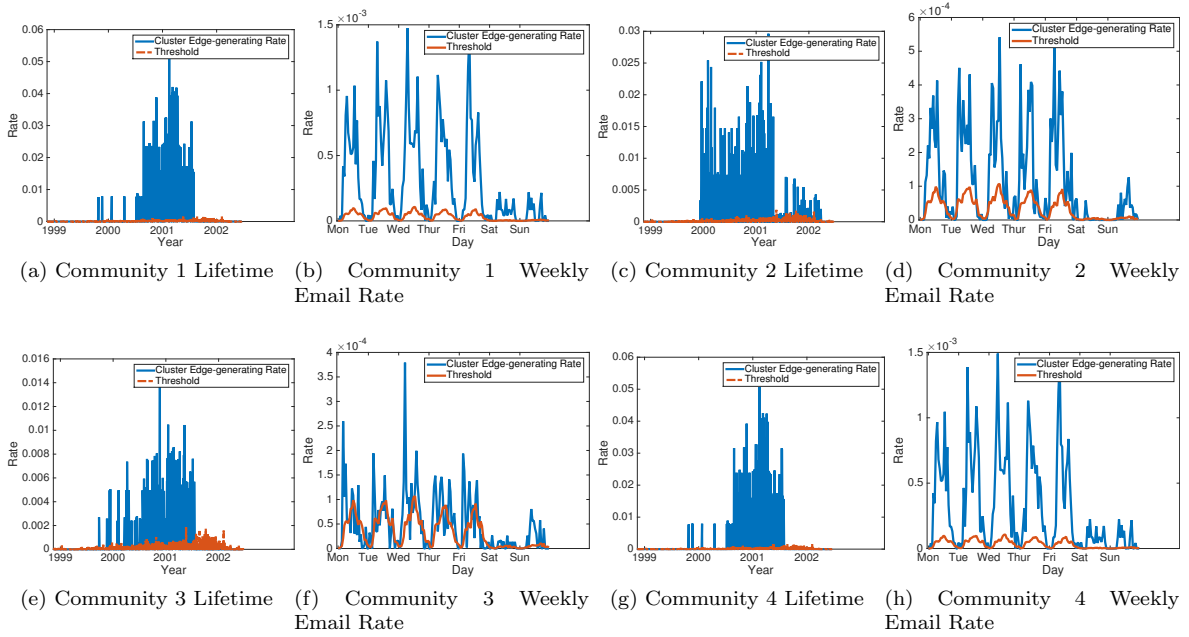

Figure 10: Communities 1-4 in Enron Email Network

Table 3: Top Active Person in Enron Community 15

| Name | Position | Role |
|---|---|---|
| Jeffrey Shankman | President | Enron Global Mkts |
| Danny McCarty | Vice President | |
| David Delainey | CEO | Enron North America and Enron Enery Services |
| Mark Haedicke | Managing Director | Legal Department |
| Greg Whalley | President | |
| James Derrick | In House Lawyer | |
| Vince Kaminski | Manager | Risk Management Head |

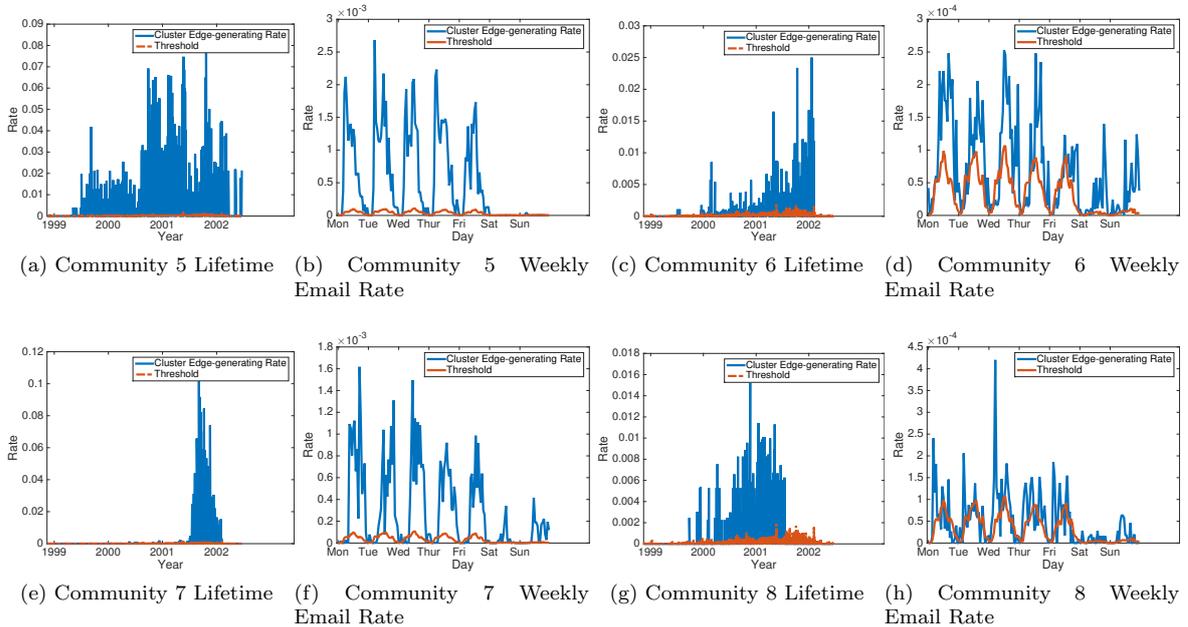

Figure 11: Communities 5-8 in Enron Email Network

### B.3 Effect of Small Rank

The rank of the decomposition, $R$, affects the error (Eq (3)), the uniqueness of the result [11] and the number of components. However, it is still unclear how it affects the meaning of components we can interpret for clusters. In the following experiments, we begin with the first and second data set to show how $R$ affects the decomposition result.

In this subsection, we show that the strengths of community structures could cause a small-rank PARAFAC decomposition to combine multiple clusters into one component.

We define the *norm of a cluster* $C_m^{(r)}$ to measure the strength of its community structure:

$$\text{norm}(C_m^{(r)}) = \sqrt{\sum_{\{i,j,k \mid i,j \in C_m^{(r)}, t_{kr} > 0\}} \mathcal{X}_{ijk}^2} \qquad (17)$$

where $t_{kr}$ is the value of $\mathcal{T}_r$ at the $m$th time step in the $r$th component. According to this definition, $\text{norm}(C_m^{(r)})$ is affected by the similarity $\mathcal{X}_{ijm}$ as well as the size of $C_m^{(r)}$. As we will see later, $\text{norm}(C_m^{(r)})$ is highly related to the conditions under which TD eliminates weak clusters or combines them into a component.

We choose $R = 1$ and apply PARAFAC on to the tensor from the first synthetic data set. To identify conditions where low-rank TD eliminates a cluster or combines clusters into one component, we specifically normalize elements $a_{i1} \in A_1$ to the range $[0, 1]$ because this allow us to qualify how likely node $i$ belongs to the component. (These normalized values are only for illustration purpose and should not be used in Eq (4))). Fig 13(a, b) shows the plots of column vectors $A_1 = [a_{1,1}, a_{2,1}, \ldots, a_{100,1}]$ and $\mathcal{T}_1 = [t_{1,1}, \ldots, t_{100,1}]$. Note that $a_{i1}$ has the same value one for node $v_i \in$ clique $G_1$ and $a_{j1}$ has the same value around 0.98 for node $v_j \in$ clique $G_2$. A $K$-mean clustering on $A_1$ generates 2 clusters, $C_1^{(1)}$ and $C_2^{(1)}$, which represent the two cliques. We use $g_1$ to represent the value of $a_{i,1}$ for node $v_i \in C_1^{(1)}$ and $g_2$ to represent the value of $a_{j,1}$ for $v_j \in C_2^{(1)}$. The difference $d = g_1 - g_2 = 0.02$ is not large, suggesting that TD has combined the two cliques into $A_1$. As a result, there is error in the decomposition(Eq (4): $\mathcal{X}_{ijk} = 0$ but $\lambda_1 a_{i1} b_{j1} t_{k1} = 0.98 t_{k1} > 0$ for $v_i \in G_1, v_j \in G_2$.

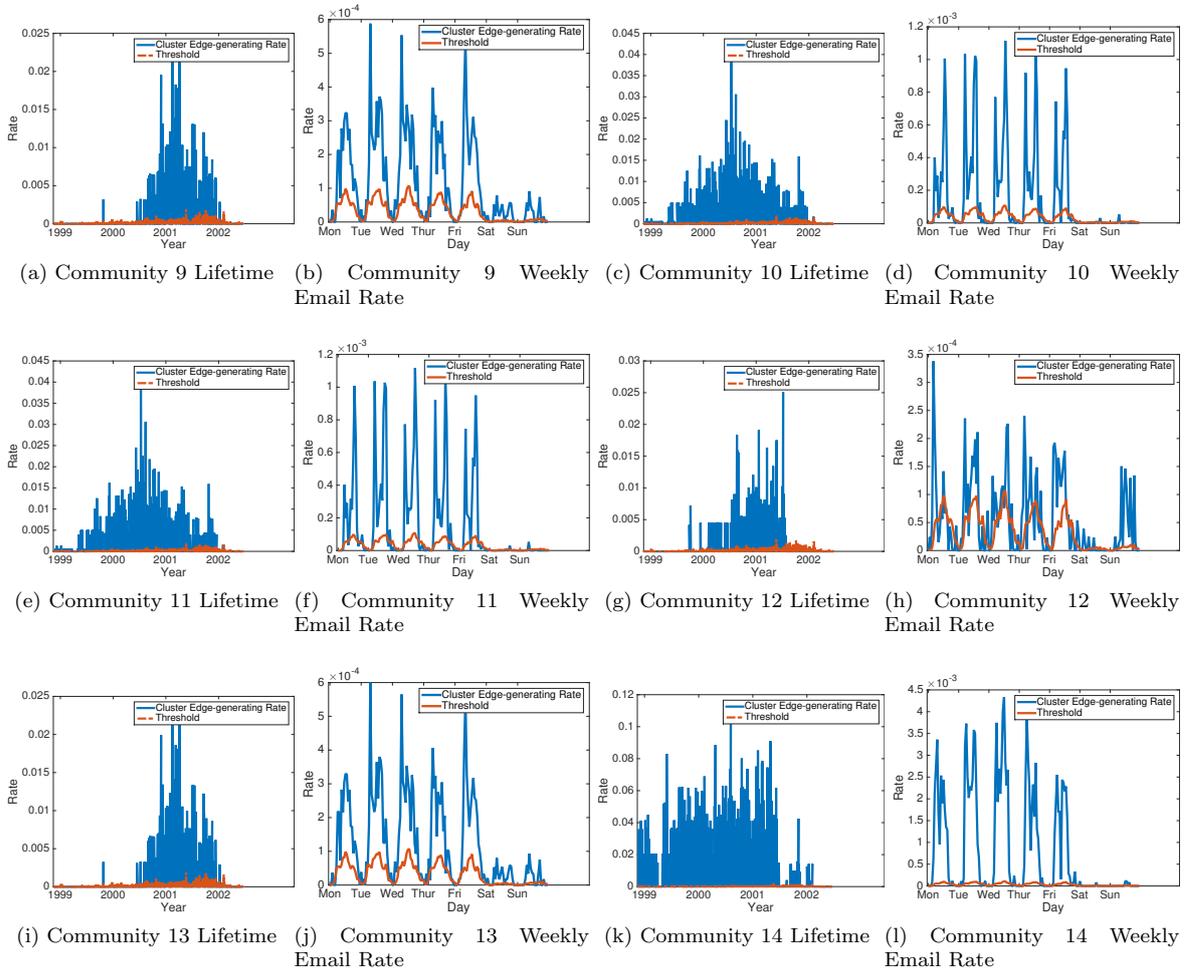

Figure 12: Communities 9-14 in Enron Email Network

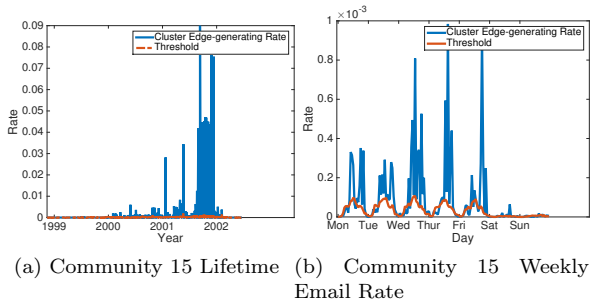

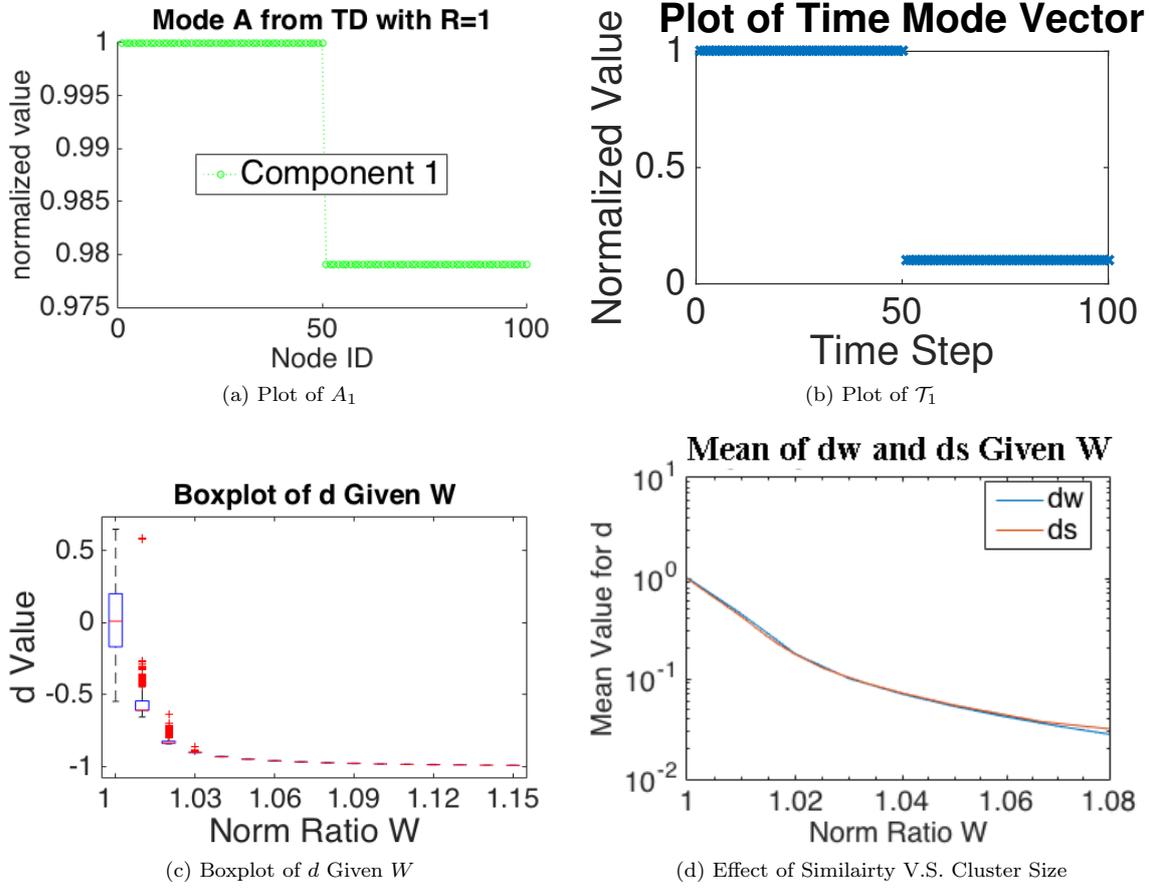

Figure 13: Effect of small $R$ on Mode A:(a) two clusters combined into $A_1$;(b) Plot of time mode $\mathcal{T}_1$; (c) $d$ decreases exponentially with $W$; (d) change of edge weight or cluster size has same effect on $d$

The value $d = g_1 - g_2 \in [-1, 1]$ shows the difference between the two cliques in the component: if $d \approx 0$, the two clusters are equally important in the component; if $d \approx 1$ or $d \approx -1$, we infer that one of the cliques is eliminated. We discover that the cluster *norms ratio* between $C_1^{(1)}$ and $C_2^{(1)}$, $W = norm(C_2^{(1)})/norm(C_1^{(1)})$, is highly related to $d$. We noted after Eq (17) that $norm(C_m^{(r)})$ depends on $\mathcal{X}_{ijk}$ and the cluster size.

We first investigate how TD behaves if cliques $G_1$ and $G_1$ are different only in $\mathcal{X}_{ijk}$, the similarity of nodes within themselves. We add different scalar value $a = 1, 1.01, \cdots, 1.15$ to $\mathcal{X}_{ijk}$, for $v_i, v_j \in G_2$, to generate new tensors and thus make $W$ change accordingly. Note that the result of a TD is not unique [25], for each $W$ value, we ran TD for 1000 times. Fig 13(c) shows boxplots of $d$ as a function of $W$. The left most boxplot for $d$ given $W = 1$ shows that $d$ falls within [-0.5 0.5] most of the time, suggesting that TD combines the two cliques given that they have the same *cluster norm*. However, as $W$, the average $\bar{d}_w$ decreases exponentially. Since $\bar{d}_w >= -1$, we model $\bar{d}_w = a * \exp(bW) + 1$ and perform a regression, obtaining $a = 1.69e + 35, b = -81.11$. When $W > 1.06$, $\bar{d}_w$ is almost $-1$, meaning that $C_2^{(1)}$ dominates the component but $C_1^{(1)}$ is eliminated.

We also investigate the effect of clique size on TD decomposition, and compare it to that caused by node similarities. We increase the size of clique $C_2^{(1)}$ to $50, 51, \cdots, 100$ and obtain 50 tensors. For each tensor, we obtain two clusters and calculate $W$. We define the average $d$ as $\bar{d}_s$. Figure 13(d) compare the log-plot of $\bar{d}_s, \bar{d}_w$ with $W \in [1, 1.08]$. There is little difference between the two line, suggesting that as long as the norm ratio $W$ is the same, TD based clustering exhibits similar behavior regardless of the difference in $\mathcal{X}_{ijk}$ or cluster size.

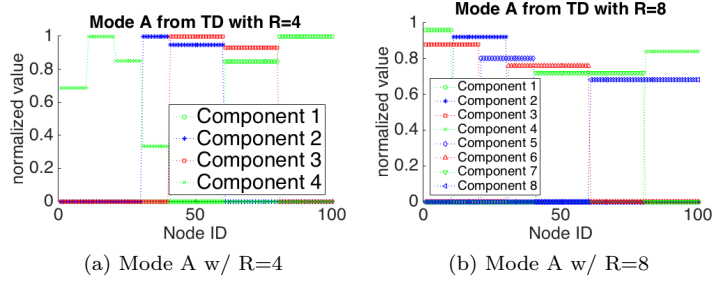

(a) Mode A w/ R=4  (b) Mode A w/ R=8

Figure 14: Plot of Mode A: (a) PARAFAC with small $R$ combines clusters with common nodes into one component; (b) PARAFAC with proper $R$ put each cluster in a separate component.

We use the second data set to investigate the condition when multiple clusters are combined into one component. We apply PARAFAC decomposition with R = 4, 8; results are shown in Figure 14. When R=4 (Fig 14(a)), clusters with common nodes are combined into one component. For example, component 4 plotted with x" represents a mixture of cl1, cl2, cl5 and cl6. These cliques are hard to distinguish because small R causes information loss, although subgroups can be detected. When R=8, each component contains only one clique. Fig 14(b) shows the plots of $A_r$, where $r = 1, \cdots, 8$. The value of $A_r$ is normalized to $[0, 1 - 0.04r]$ for clear illustration.

In summary, the effect of $R$ on clustering on TD components is:

- If $R$ is smaller than the number of clusters in the dynamic graph, independent clusters with similar norm value, or clusters with common nodes tend to combine into one component. Cluster detection may become difficult in this case. Meanwhile, clusters with small norm are likely to be eliminated.

- If $R$ is large, TD component tend to contain fewer clusters. Ground truth clusters are more easily detected.

# References

[1] Evrim Acar, Daniel M. Dunlavy, and Tamara G. Kolda. A scalable optimization approach for fitting canonical tensor decompositions. *Journal of Chemometrics*, 25(2):67–86, February 2011.

[2] Evrim Acar, Daniel M Dunlavy, Tamara G Kolda, and Morten Mrup. Scalable tensor factorizations with missing data. In *SDM*, pages 701–712. SIAM, 2010.

[3] Amr Ahmed and Eric P Xing. Dynamic non-parametric mixture models and the recurrent chinese restaurant process: with applications to evolutionary clustering. In *SDM*, pages 219–230. SIAM, 2008.

[4] Brett W. Bader and Tamara G. Kolda. Matlab tensor toolbox version 2.6. http://www.sandia.gov/~tgkolda/TensorToolbox/, February 2015.

[5] Rasmus Bro and Henk AL Kiers. A new efficient method for determining the number of components in parafac models. *Journal of chemometrics*, 17(5):274–286, 2003.

[6] Deepayan Chakrabarti, Ravi Kumar, and Andrew Tomkins. Evolutionary clustering. In *Proceedings of the 12th ACM SIGKDD international conference on Knowledge discovery and data mining*, pages 554–560. ACM, 2006.

[7] Yung-Chih Chen, Elisha Rosensweig, Jim Kurose, and Don Towsley. Group detection in mobility traces. In *Proceedings of the 6th international wireless communications and mobile computing conference*, pages 875–879. ACM, 2010.

[8] Jana Diesner and Kathleen M Carley. Exploration of communication networks from the enron email corpus. In *SIAM International Conference on Data Mining: Workshop on Link Analysis, Counterterrorism and Security, Newport Beach, CA*, 2005.

[9] Wenjie Fu, Le Song, and Eric P. Xing. Dynamic mixed membership blockmodel for evolving networks. In *Proceedings of the 26th Annual International Conference on Machine Learning*, ICML '09, pages 329–336, New York, NY, USA, 2009. ACM.

[10] Laetitia Gauvin, Andr Panisson, and Ciro Cattuto. Detecting the community structure and activity patterns of temporal networks: a non-negative tensor factorization approach. *PloS one*, 9(1):e86028, 2014.

[11] Harshman and Richard A. Foundations of the parafac procedure: Models and conditions for an" explanatory" multi-modal factor analysis. *UCLA Working Papers in phonetics*, 16:1, 1970.

[12] Eamonn Keogh, Selina Chu, David Hart, and Michael Pazzani. Segmenting time series: A survey and novel approach. *Data mining in time series databases*, 57:1–22, 2004.

[13] David J Ketchen and Christopher L Shook. The application of cluster analysis in strategic management research: an analysis and critique. *Strategic management journal*, 17(6):441–458, 1996.

[14] Min-Soo Kim and Jiawei Han. A particle-and-density based evolutionary clustering method for dynamic networks. *Proceedings of the VLDB Endowment*, 2(1):622–633, 2009.

[15] Stefano Leonardi, Aris Anagnostopoulos, Jakublacki, Silvio Lattanzi, and Mohammad Mahdian. Community detection on evolving graphs. In *Advances in Neural Information Processing Systems*, pages 3522–3530, 2016.

[16] Yu-Ru Lin, Yun Chi, Shenghuo Zhu, Hari Sundaram, and Belle L Tseng. Facetnet: a framework for analyzing communities and their evolutions in dynamic networks. In *Proceedings of the 17th international conference on World Wide Web*, pages 685–694. ACM, 2008.

[17] Hing-Hao Mao, Chung-Jung Wu, Evangelos E Papalexakis, Christos Faloutsos, Kuo-Chen Lee, and Tien-Cheu Kao. Malspot: Multi2 malicious network behavior patterns analysis. In *Advances in Knowledge Discovery and Data Mining*, pages 1–14. Springer, 2014.

[18] Mark EJ Newman. Modularity and community structure in networks. *Proceedings of the National Academy of Sciences*, 103(23):8577–8582, 2006.

[19] Mark EJ Newman. Modularity and community structure in networks. *Proceedings of the national academy of sciences*, 103(23):8577–8582, 2006.

[20] Andrew Y Ng, Michael I Jordan, Yair Weiss, et al. On spectral clustering: Analysis and an algorithm. In *NIPS*, volume 14, pages 849–856, 2001.

[21] Evangelos E Papalexakis, Konstantinos Pelechrinis, and Christos Faloutsos. Location based social network analysis using tensors and signal processing tools. In *Computational Advances in Multi-Sensor Adaptive Processing (CAMSAP), 2015 IEEE 6th International Workshop on*, pages 93–96. IEEE, 2015.

[22] Carey E Priebe, John M Conroy, David J Marchette, and Youngser Park. Scan statistics on enron graphs. *Computational and Mathematical Organization Theory*, 11(3):229–247, 2005.

[23] Peter J Rousseeuw. Silhouettes: a graphical aid to the interpretation and validation of cluster analysis. *Journal of computational and applied mathematics*, 20:53–65, 1987.

[24] Richard Sands and Forrest W Young. Component models for three-way data: An alternating least squares algorithm with optimal scaling features. *Psychometrika*, 45(1):39–67, 1980.

[25] Alwin Stegeman and Nicholas D Sidiropoulos. On kruskals uniqueness condition for the candecomp/parafac decomposition. *Linear Algebra and its applications*, 420(2):540–552, 2007.

[26] Lei Tang, Huan Liu, Jianping Zhang, and Zohreh Nazeri. Community evolution in dynamic multi-mode networks. In *Proceedings of the 14th ACM SIGKDD international conference on Knowledge discovery and data mining*, pages 677–685. ACM, 2008.

[27] PC URC. Network analysis with the enron email corpus. *Journal of Statistics Education*, 23(2), 2015.

[28] Kevin S Xu, Mark Kliger, and Alfred O Hero. Evolutionary spectral clustering with adaptive forgetting factor. In *Acoustics Speech and Signal Processing (ICASSP), 2010 IEEE International Conference on*, pages 2174–2177. IEEE, 2010.

[29] Yangyang Xu and Wotao Yin. A block coordinate descent method for regularized multiconvex optimization with applications to nonnegative tensor factorization and completion. *SIAM Journal on imaging sciences*, 6(3):1758–1789, 2013.

[30] Tianbao Yang, Yun Chi, Shenghuo Zhu, Yihong Gong, and Rong Jin. Detecting communities and their evolutions in dynamic social networks: a bayesian approach. *Machine learning*, 82(2):157–189, 2011.